\documentclass[11pt,a4paper]{amsart}
\usepackage{amsmath,amssymb,latexsym,amsthm,enumerate}
\usepackage{graphicx} 
\usepackage{hyperref} 
\usepackage{xcolor} 
\usepackage{psfrag} 

\usepackage{verbatim}

\theoremstyle{plain}
\newtheorem{theorem}{Theorem}[section]
\newtheorem{proposition}[theorem]{Proposition}
\newtheorem{lemma}[theorem]{Lemma}
\newtheorem{corollary}[theorem]{Corollary}
\theoremstyle{definition}

\newtheorem{example}[theorem]{Example}

\theoremstyle{remark}

\newtheorem*{remark}{Remark}
\newtheorem*{remarks}{Remarks}

\newcounter{numpar}[section]

\newcommand*{\wt}{\widetilde}
\newcommand*{\ol}{\overline}

\newcommand*{\eps}{\varepsilon}

\newcommand*{\cF}{\mathcal F}

\newcommand*{\cM}{{\mathcal M}}

\newcommand*{\bbN}{\mathbb N}

\newcommand*{\bbR}{\mathbb R}

\newcommand*{\EE}{\mathsf E}
\newcommand*{\PP}{\mathsf P}
\newcommand*{\QQ}{\mathsf Q}

\newcommand*{\tPP}{{\wt\PP}}
\newcommand*{\tQQ}{{\wt\QQ}}

\newcommand*{\LlocJ}{{L^1_{\mathrm{loc}}(J)}}
\newcommand*{\Llocr}{{L^1_{\mathrm{loc}}(r-)}}
\newcommand*{\Llocl}{{L^1_{\mathrm{loc}}(l+)}}
\newcommand*{\Llocinfty}{{L^1_{\mathrm{loc}}(\infty-)}}
\newcommand*{\LlocZ}{{L^1_{\mathrm{loc}}(0+)}}

\begin{document}
\title[Arbitrage in Diffusion Models]{Deterministic criteria for the absence of arbitrage in one-dimensional diffusion models}

\author{Aleksandar Mijatovi\'{c}}
\address{Department of Mathematics, Imperial College London, UK}
\email{a.mijatovic@imperial.ac.uk}

\author{Mikhail Urusov}
\address{Institute of Mathematical Finance, Ulm University, Germany}
\email{mikhail.urusov@uni-ulm.de}

\thanks{We are grateful to Peter Bank, Nicholas Bingham,
Mark Davis, Yuri Kabanov, Ioannis Karatzas, Walter Schachermayer,
Martin Schweizer and the anonymous referees for valuable suggestions
which led to numerous improvements of the original manuscript.
This paper was written while the second author was a postdoc
in the Deutsche Bank Quantitative Products Laboratory, Berlin.}



\begin{abstract}
We obtain a deterministic characterisation
of the \emph{no free lunch with vanishing risk},
the \emph{no generalised arbitrage}
and the \emph{no relative arbitrage}
conditions in the one-dimensional diffusion setting
and examine how these notions of no-arbitrage
relate to each other.
\end{abstract}

\maketitle

\section{Introduction}
\label{I}
In this paper we consider a market that consists of a money market account
and a risky asset whose discounted price is given by a nonnegative process $Y$
satisfying the SDE
\begin{equation}
\label{eq:i1}
dY_t=\mu(Y_t)\,dt+\sigma(Y_t)\,dW_t,\quad Y_0=x_0>0.
\end{equation}
We are interested in the notions of \emph{free lunch with vanishing risk}
(see Delbaen and Schachermayer \cite{DelbaenSchachermayer:94} and~\cite{DelbaenSchachermayer:98}),
\emph{generalised arbitrage} (see Sin~\cite{Sin:96}, Yan~\cite{Yan:98} and Cherny~\cite{Cherny:07})
and \emph{relative arbitrage} (see Fernholz and Karatzas~\cite{FernholzKaratzas:09}).
In what follows we use the acronyms FLVR, GA and RA for the notions above
and the acronyms NFLVR, NGA and NRA for the corresponding types of no-arbitrage.

The notion of FLVR was introduced by Delbaen and Schachermayer~\cite{DelbaenSchachermayer:94}
(see also~\cite{DelbaenSchachermayer:98} and the monograph~\cite{DelbaenSchachermayer:06})
and is by now a classical notion of arbitrage in continuous-time models.
We recall the definition in Section~\ref{FLVR}.
The notion of GA was introduced independently and in different ways
in Sin~\cite{Sin:96}, Yan~\cite{Yan:98} and Cherny~\cite{Cherny:07}
(the term \emph{generalised arbitrage} comes from~\cite{Cherny:07}).
In continuous time the approaches in \cite{Sin:96}, \cite{Yan:98} and~\cite{Cherny:07}
provide a new look at no-arbitrage and the valuation of derivatives.
We recall the definition in Section~\ref{GA}.
The requirement of NGA is stronger than that of NFLVR,
and the difference comes loosely speaking from the fact
that a wider set of admissible strategies is considered when defining NGA.
To obtain an intuitive understanding of the difference between these two notions
consider for example the discounted price process $(Y_t)_{t\in[0,1]}$
that is a local martingale with $Y_0=1$ and $Y_1=0$ (hence not a martingale).
There exists GA in this model and it consists of selling the asset short
at time~$0$ and buying it back at time~$1$.
However this model satisfies the NFLVR condition: the strategy above is non-admissible
in the framework of Delbaen and Schachermayer because its wealth process
$(Y_0-Y_t)_{t\in[0,1]}$ is unbounded from below.

The notion of RA was introduced within the framework of stochastic portfolio theory (SPT),
proposed in recent years as a tool for analysing the observed phenomena
in the equity markets and optimizing portfolio allocation in the long run
(see Fernholz~\cite{Fernholz:02} and Fernholz and Karatzas~\cite{FernholzKaratzas:09}).
From this viewpoint SPT resembles the benchmark approach in finance
(see Platen and Heath~\cite{PlatenHeath:06}).
SPT is a descriptive theory that descends from the classical portfolio theory
of Harry Markowitz and in many ways departs from the well-known paradigm
of dynamic asset pricing. Informally, there is \emph{arbitrage relative to the market}
(or simply \emph{relative arbitrage}, RA) if there exists an investment strategy
that beats the market portfolio (for more details see e.g.
Fernholz, Karatzas and Kardaras~\cite{FernholzKaratzasKardaras:05},
Fernholz and Karatzas~\cite{FernholzKaratzas:09} and Ruf~\cite{Ruf:09}).
This reduces in the one-dimensional setting considered in this paper
to the existence of an investment strategy that beats the stock~$Y$.
It is therefore especially interesting to examine the relation between RA and FLVR,
since the latter notion is based on the related but different idea of the existence
of an investment strategy that beats the money market account.

The main contribution of the present paper is that it gives
deterministic necessary and sufficient conditions
for the absence of FLVR, GA and RA in the diffusion model~\eqref{eq:i1},
all of which are expressed in terms of the drift
$\mu$ and the volatility~$\sigma$.
The diffusion setting considered here is quite general as the coefficients
of SDE~\eqref{eq:i1}
are Borel measurable functions that are
only required to satisfy a weak local integrability assumption
and the process
$Y$
is allowed to reach zero in finite time.
Deterministic characterisation of no-arbitrage conditions is, to 
our knowledge, not common in the literature. The only instance
known to us is the work of Delbaen and Shirakawa~\cite{DelbaenShirakawa:02b}
where a necessary and sufficient condition for NFLVR is developed under
more restrictive assumptions on the underlying diffusion. In fact 
Theorem~\ref{FLVR1} in this paper can be viewed as a generalisation 
of the characterisation result
in~\cite{DelbaenShirakawa:02b}
(see the remark following Theorem~\ref{FLVR1} for details).

One of the ingredients of the proof of Theorem~\ref{FLVR1}
is the central theorem in~\cite{MijatovicUrusov:09a},
which characterises the martingale property of 
certain stochastic exponentials. It is important
to stress however that Theorem~\ref{FLVR1},
which states the deterministic necessary and sufficient condition 
for NFLVR,
is not a simple consequence of the characterisation of the
martingale property given in~\cite{MijatovicUrusov:09a}.
There are three reasons for this. The first is that the 
characterisation result in~\cite{MijatovicUrusov:09a}
only applies under assumption~(8) in~\cite{MijatovicUrusov:09a},
which when translated into the setting of the present paper 
corresponds to condition~\eqref{sna2}.
Theorem~\ref{FLVR1} applies without assuming~\eqref{sna2}.
In fact
the deterministic necessary and sufficient condition
for NFLVR given in this theorem 
shows that property~\eqref{sna2}
plays a key role in determining whether a diffusion model~\eqref{eq:i1}
satisfies the NFLVR condition.
The second reason is that even in the case where assumption~\eqref{sna2}
holds, the main result of~\cite{MijatovicUrusov:09a} 
implies only the absolute continuity of the local martingale measure
with respect to the original probability measures.
The equivalence of measures can only be obtained as a consequence of 
Theorems~\ref{MRA3} and~\ref{MRA4},
which are established in the present paper.
The third reason is that in the general continuous semimartingale setting 
an equivalent local martingale measure can exist with a
density process different from the stochastic exponential 
of the Girsanov type
(an example of such a model is given 
in~\cite{DelbaenSchachermayer:98a}),
which is in our setting given by the process $Z$
(see~\eqref{mra1} for definition)
studied in~\cite{MijatovicUrusov:09a}.

The related question of a (non-deterministic) characterisation
of NFLVR in the class of models given by
It\^o processes was studied by Lyasoff~\cite{Lyasoff:08}.
In such a market model a pathwise square integrability condition
(assumption~(1.1) in~\cite{Lyasoff:08})
on the market price of risk process is natural
and furthermore has to be assumed for the model to have desirable properties
(e.g. if (1.1) in~\cite{Lyasoff:08} does not hold the model allows arbitrage).
However such a condition is difficult to verify
if the price process is a solution of SDE~\eqref{eq:i1},
since the market price of risk is in this case
implicitly determined by the coefficients of the SDE,
which only satisfy mild local integrability assumptions.
In fact a solution of SDE~\eqref{eq:i1} can exist and
be unique while the corresponding market price of risk
does not possess the required property.
Moreover the answer in~\cite{Lyasoff:08}
is given in a form that is very different from ours.

Once the deterministic necessary and sufficient conditions
for the absence of various types of arbitrage
have been established, we apply them
to  examine how these notions relate to each other.
When studying the various notions of arbitrage
we suppose that $Y$ does not explode at~$\infty$
but may reach zero in finite time.
The assumption of non-explosion at $\infty$ is natural for a stock price process.
Although it may seem natural also to exclude the possibility of explosion at zero,
we do not do so as such behaviour is exhibited by some models
considered in the literature (e.g. the CEV model).
Let the process $Z$ be the candidate for the density of the equivalent local martingale measure
in our model.
As we shall see, if the diffusion
$Y$ reaches zero at a finite time,
the process
$Z$ may also reach zero, 
however it may also happen that $Z$ remains strictly positive.
As mentioned above
in order to obtain a sufficient condition for NFLVR
(i.e. prove that the local martingale measure is equivalent,
not just absolutely continuous)
we will need to analyse when $Z$ reaches zero at a finite time.
This analysis is carried out in Section~\ref{MR}
in a slightly more general setting, which may be of interest
also in other contexts.
Section~\ref{SNA} presents the deterministic characterisation of
NFLVR, NGA and NRA in model~\eqref{eq:i1}.
In Subsection~\ref{SNAC} we prove that in general
NFLVR and NRA neither imply nor exclude each other, and
that in the class of models given by~\eqref{eq:i1},
where all three notions can be defined simultaneously,
the relationship
\begin{equation*}
\text{NGA}\qquad\Longleftrightarrow\qquad\text{NFLVR}\quad\&\quad\text{NRA}
\end{equation*}
holds. The proofs of the characterisation theorems of Sections~\ref{MR} and~\ref{FLVR}
are given in Section~\ref{PFLVR}.

\section{Is the candidate for the density process strictly positive?}
\label{MR}
We consider the state space $J=(l,r)$, $-\infty\le l<r\le\infty$
and a $J$-valued diffusion $Y=(Y_t)_{t\in[0,\infty)}$
on some filtered probability space
$(\Omega,\cF,(\cF_t)_{t\in[0,\infty)},\PP)$
driven by the SDE
\begin{equation}
\label{mr1}
dY_t=\mu(Y_t)\,dt+\sigma(Y_t)\,dW_t,\quad Y_0=x_0\in J,
\end{equation}
where $W$ is an $(\cF_t)$-Brownian motion and $\mu,\sigma\colon J\to\bbR$
are Borel functions  satisfying the Engelbert--Schmidt conditions
\begin{gather}
\label{mr2} 
\sigma(x)\ne0\;\;\forall x\in J,\\
\label{mr3}
\frac1{\sigma^2},\;\frac\mu{\sigma^2}\in\LlocJ.
\end{gather}
$\LlocJ$ denotes the class of locally integrable functions,
i.e. the functions $J\to\bbR$ that are integrable on compact subsets of~$J$.
Under conditions~\eqref{mr2} and~\eqref{mr3} SDE~\eqref{mr1}
has a unique in law (possibly explosive) weak solution
(see~\cite{EngelbertSchmidt:85}, \cite{EngelbertSchmidt:91},
or \cite[Ch.~5, Th.~5.15]{KaratzasShreve:91}).
By $\zeta$ we denote the explosion time of~$Y$.
In the case $\PP(\zeta<\infty)>0$ we need to specify
the behaviour of $Y$ after explosion.
In what follows we assume that the solution
$Y$
on the set $\{\zeta<\infty\}$ 
stays after $\zeta$ at the boundary point of
$J$
at which it explodes,
i.e. $l$ and $r$ become absorbing boundaries.
We will use the following terminology:

\emph{$Y$ explodes at $r$} means $\PP(\zeta<\infty,\lim_{t\uparrow\zeta}Y_t=r)>0$;

\emph{$Y$ explodes at $l$} is understood in  a similar way.

\noindent
The Engelbert--Schmidt conditions are reasonable
weak assumptions. For instance, they are satisfied
if $\mu$ is locally bounded on $J$
and $\sigma$ is locally bounded away from zero on~$J$.
Finally, let us note that we assume neither that $(\cF_t)$ is generated by $W$
nor that $(\cF_t)$ is generated by~$Y$.

In this section we consider the stochastic exponential
\begin{equation}
\label{mra1}
Z_t=\exp\left\{\int_0^{t\wedge\zeta}b(Y_u)\,dW_u
-\frac12\int_0^{t\wedge\zeta}b^2(Y_u)\,du\right\},
\quad t\in[0,\infty),
\end{equation}
where we set $Z_t:=0$ for $t\ge\zeta$ on
$\{\zeta<\infty,\int_0^\zeta b^2(Y_u)\,du=\infty\}$.
In what follows we assume that
$b$ is a Borel function $J\to\bbR$ satisfying
\begin{equation}
\label{mra2}
\frac{b^2}{\sigma^2}\in\LlocJ.
\end{equation}
In particular, $b$ could be an arbitrary locally bounded function on~$J$.
Using the occupation times formula it is easy to
show that condition~\eqref{mra2} is equivalent to
\begin{equation}
\label{mra3}
\int_0^t b^2(Y_u)\,du<\infty \quad\PP\text{-a.s. on }\{t<\zeta\},\quad t\in[0,\infty).
\end{equation}
We need to assume condition~\eqref{mra3} to ensure that the stochastic integral
$\int_0^t b(Y_u)\,dW_u$ is well-defined on $\{t<\zeta\}$,
which is equivalent to imposing~\eqref{mra2} on the function~$b$.
Thus the defined process $Z=(Z_t)_{t\in[0,\infty)}$
is a nonnegative continuous local martingale
(continuity at time $\zeta$ on the set $\{\zeta<\infty,\int_0^\zeta b^2(Y_u)\,du=\infty\}$
follows from the Dambis--Dubins--Schwarz theorem; see~\cite[Ch.~V, Th.~1.6 and Ex.~1.18]{RevuzYor:99}).

As a nonnegative local martingale $Z$ is a supermartingale.
Hence, it has a finite limit $Z_\infty=(\PP\text{-a.s.})\lim_{t\to\infty}Z_t$.
In Theorem~\ref{MRA3} below we give a deterministic necessary and sufficient condition
for $Z$ to be strictly positive. In Theorem~\ref{MRA4} we present a deterministic
criterion for $Z_\infty>0$~$\PP$-a.s. Let us note that the condition
$Z_\infty>0$~$\PP$-a.s. implies strict positivity of $Z$
as, clearly, $Z$ stays at zero after it hits zero.
Finally, in Theorem~\ref{MRA5} we provide a criterion for $Z_\infty=0$~$\PP$-a.s.

Before we formulate these results let us introduce some notation.
Let $\ol J:=[l,r]$.
Let us fix an arbitrary $c\in J$ and set
\begin{align}
\label{mra8}
\rho(x)&=\exp\left\{-\int_c^x\frac{2\mu}{\sigma^2}(y)\,dy\right\},
\quad x\in J,\\
\label{mra9}
s(x)&=\int_c^x\rho(y)\,dy,\quad x\in\ol J.
\end{align}
Note that $s$ is the scale function of diffusion~\eqref{mr1}.
By $\Llocr$ we denote the class of Borel functions $f\colon J\to\bbR$
such that $\int_x^r |f(y)|\,dy<\infty$ for some $x\in J$.
Similarly we introduce the notation $\Llocl$.

Let us recall that the process $Y$ explodes at the boundary point $r$ if and only if
\begin{equation}
\label{mra21}
s(r)<\infty\text{ and }\frac{s(r)-s}{\rho\sigma^2}\in\Llocr.
\end{equation}
This is Feller's test for explosions
(see e.g. \cite[Sec.~4.1]{ChernyEngelbert:05}
or \cite[Ch.~5, Th.~5.29]{KaratzasShreve:91}).
Similarly, $Y$ explodes at the boundary point $l$ if and only if
\begin{equation}
\label{mra23}
s(l)>-\infty\text{ and }\frac{s-s(l)}{\rho\sigma^2}\in\Llocl.
\end{equation}
We say that the endpoint $r$ of $J$ is \emph{good} if
\begin{equation}
\label{mra12}
s(r)<\infty\text{ and }\frac{(s(r)-s)b^2}{\rho\sigma^2}\in\Llocr.
\end{equation}
We say that the endpoint $l$ of $J$ is \emph{good} if
\begin{equation}
\label{mra14}
s(l)>-\infty\text{ and }\frac{(s-s(l))b^2}{\rho\sigma^2}\in\Llocl.
\end{equation}
If $l$ or~$r$ is not good, we call it \emph{bad}.

In the following theorem let $T\in(0,\infty)$
be a fixed finite time.

\begin{theorem}
\label{MRA3}
Let the functions $\mu$, $\sigma$ and $b$ satisfy conditions
\eqref{mr2}, \eqref{mr3} and~\eqref{mra2},
and $Y$ be a (possibly explosive) solution of SDE~\eqref{mr1}.
Then we have $Z_T>0$~$\PP$-a.s.
if and only if at least one of the conditions (a)--(b) below is satisfied
AND at least one of the conditions (c)--(d) below is satisfied:

(a) $Y$ does not explode at~$r$, i.e. \eqref{mra21} is not satisfied;

(b) $r$ is good, i.e. \eqref{mra12} is satisfied;

(c) $Y$ does not explode at~$l$, i.e. \eqref{mra23} is not satisfied;

(d) $l$ is good, i.e. \eqref{mra14} is satisfied.
\end{theorem}

\begin{remark}
Clearly, the process $Z$ stays at zero after it hits zero.
Therefore, the condition $Z_T>0$ $\PP$-a.s. is equivalent to the condition
that the process $(Z_t)_{t\in[0,T]}$ is $\PP$-a.s. strictly positive.
Furthermore, since none of conditions (a)--(d) of Theorem~\ref{MRA3} involve~$T$,
the criterion in Theorem~\ref{MRA3} is also the criterion for ascertaining that
the process $(Z_t)_{t\in[0,\infty)}$ is $\PP$-a.s. strictly positive.
\end{remark}

\begin{theorem}
\label{MRA4}
Under the assumptions of Theorem~\ref{MRA3} we have $Z_\infty>0$~$\PP$-a.s.
if and only if at least one of the conditions (I)--(IV) below is satisfied:

(I) $b=0$ a.e. on $J$ with respect to the Lebesgue measure;

(II) $r$ is good and $s(l)=-\infty$;

(III) $l$ is good and $s(r)=\infty$;

(IV) $l$ and $r$ are good.
\end{theorem}

\begin{remark}
Condition~(I) cannot be omitted here.
Indeed, if $J=\bbR$, $b\equiv0$ and $Y=W$,
then $Z\equiv1$, so $Z_\infty>0$~a.s.,
but none of conditions (II), (III) and~(IV) hold
because $s(-\infty)=-\infty$ and $s(\infty)=\infty$
(and hence neither endpoint is good).
\end{remark}

\begin{theorem}
\label{MRA5}
Under the assumptions of Theorem~\ref{MRA3} we have $Z_\infty=0$~$\PP$-a.s.
if and only if both conditions (i) and~(ii) below are satisfied:

(i) $b$ is not identically zero (with respect to the Lebesgue measure);

(ii) $l$ and $r$ are bad.
\end{theorem}

Condition~(i) cannot be omitted here
(see the remark following Theorem~\ref{MRA4}).

The proofs of Theorems~\ref{MRA3}, \ref{MRA4} and~\ref{MRA5}
are based on 
the notion of separating time 
and will be given in Section~\ref{PFLVR}.

\medskip
To apply the theorems above we need to check in specific situations
whether the endpoints $l$ and $r$ are good.
Below we quote two remarks, proved in~\cite{MijatovicUrusov:09a},
that can facilitate these checks and will be used in the sequel.
Let us consider an auxiliary $J$-valued diffusion $\wt Y$ governed by the SDE
\begin{equation}
\label{mra16}
d\wt Y_t=(\mu+b\sigma)(\wt Y_t)\,dt+\sigma(\wt Y_t)\,d\wt W_t,\quad \wt Y_0=x_0,
\end{equation}
on some probability space
$(\wt\Omega,\wt\cF,(\wt\cF_t)_{t\in[0,\infty)},\tPP)$.
SDE~\eqref{mra16} has a unique in law (possibly explosive)
weak solution because the Engelbert--Schmidt conditions
\eqref{mr2} and~\eqref{mr3} are satisfied
for the coefficients $\mu+b\sigma$ and $\sigma$
(note that $b/\sigma\in\LlocJ$ holds due to~\eqref{mra2}).
As in the case of SDE~\eqref{mr1} we denote the explosion time of $\wt Y$
by $\wt\zeta$ and apply the same convention as before: 
on the set $\{\wt\zeta<\infty\}$ the solution $\wt Y$
stays after $\wt\zeta$ at the boundary point at which it explodes.
Similarly to the notations $s$ and $\rho$
let us introduce the notation $\wt s$ for the scale function of diffusion~\eqref{mra16}
and $\wt\rho$ for the derivative of~$\wt s$.

\begin{remarks}
(i) Under condition~\eqref{mra2} the endpoint $r$ of $J$ is good if and only if
\begin{equation}
\label{mra13}
\wt s(r)<\infty\text{ and }\frac{(\wt s(r)-\wt s)b^2}{\wt\rho\sigma^2}\in\Llocr.
\end{equation}
Under condition~\eqref{mra2} the endpoint $l$ of $J$ is good if and only if
\begin{equation}
\label{mra15}
\wt s(l)>-\infty\text{ and }\frac{(\wt s-\wt s(l))b^2}{\wt\rho\sigma^2}\in\Llocl.
\end{equation}
When the auxiliary diffusion~\eqref{mra16} has a simpler form than the initial diffusion~\eqref{mr1},
it may be easier to check \eqref{mra13} and~\eqref{mra15}
rather than \eqref{mra12} and~\eqref{mra14}.

\smallskip
(ii) The endpoint $r$ (resp.~$l$) is bad
whenever one of the processes $Y$ and $\wt Y$ explodes at $r$ (resp. at~$l$)
and the other does not.
This is helpful because one can sometimes immediately see that,
for example, $Y$ does not explode at $r$ while $\wt Y$ does.
In such a case one concludes that $r$ is bad without
having to check either~\eqref{mra12} or~\eqref{mra13}.
\end{remarks}

In this paper we will need to apply Theorem~2.1 from~\cite{MijatovicUrusov:09a} several times.
Each time some work needs to be done to check certain conditions in that theorem.
For the reader's convenience we quote that result below.

\begin{theorem}
\label{th:MRold}
Under the assumptions of Theorem~\ref{MRA3}
the process $Z$ is a martingale
if and only if at least one of the conditions (a') and~(b) is satisfied
AND at least one of the conditions (c') and~(d) is satisfied,
where conditions (b) and~(d) are those from Theorem~\ref{MRA3}
and conditions (a') and~(c') are given below:

(a') $\wt Y$ does not explode at~$r$;

(c') $\wt Y$ does not explode at~$l$.
\end{theorem}

\begin{example}
\label{ex:EAA2}
In this example we demonstrate how the theorems of this section
can be applied in practice. Consider a generalised constant elasticity 
of variance (CEV) process that is given by the SDE
\begin{equation}
\label{ea1}
dY_t=\mu_0 Y_t^\alpha dt+\sigma_0 Y_t^\beta dW_t,\quad Y_0=x_0\in J:=(0,\infty),\quad \alpha,\beta\in\bbR,\>\mu_0\in\bbR\setminus\{0\},\>\sigma_0>0.
\end{equation}
Note that the drift and volatility functions in~\eqref{ea1}
satisfy the conditions in~\eqref{mr2} and \eqref{mr3}.
We are interested in the stochastic exponential
\begin{equation}
\label{eq:eaa2}
Z_t=\exp\left\{
-\frac{\mu_0}{\sigma_0}\int_0^{t\wedge\zeta}Y_u^{\alpha-\beta}\,dW_u
-\frac12\frac{\mu_0^2}{\sigma_0^2}\int_0^{t\wedge\zeta}Y_u^{2\alpha-2\beta}\,du
\right\},\quad t\in[0,\infty),
\end{equation}
where we set $Z_t:=0$ for $t\ge\zeta$ on
$\{\zeta<\infty,\int_0^\zeta Y_u^{2\alpha-2\beta}\,du=\infty\}$,
which is the process of~\eqref{mra1} with
$b(x):=-\mu_0 x^{\alpha-\beta}/\sigma_0$
(clearly, \eqref{mra2} is satisfied).
Note that the auxiliary diffusion $\wt Y$,
given by~\eqref{mra16},
in this case follows the driftless SDE
$d\wt Y_t=\sigma_0 \wt Y_t^\beta d\wt W_t$, $\wt Y_0=x_0$.

We now apply the above results to determine whether the process $Z$
and its limit $Z_\infty$ are strictly positive $\PP$-a.s.
Let us note that the case $\mu_0=0$ is trivial
and therefore excluded in~\eqref{ea1}.
Since $\wt Y$ has no drift, we may take
$\wt\rho\equiv1$ and $\wt s(x)=x$.
It follows from \eqref{mra13} and~\eqref{mra15}
that $\infty$ is always a bad boundary point
and that $0$ is a good boundary point if and only if
$\alpha+1>2\beta$.
Theorem~\ref{MRA5} implies that $Z_\infty=0$~$\PP$-a.s.
if and only if $\alpha+1\le2\beta$.
Let us consider the following three cases.

Case~1: $\alpha+1<2\beta$.
A simple computation shows that $s(\infty)=\infty$,
hence $Y$ does not explode at~$\infty$.
By Theorem~\ref{MRA3}, the process $Z=(Z_t)_{t\in[0,\infty)}$
is $\PP$-a.s. strictly positive if and only if
$Y$ does not explode at~$0$.
Another simple computation yields
that the latter holds if and only if
$\mu_0>0$ or $\alpha\ge1$.

Case~2: $\alpha+1=2\beta$.
At first we find that $Y$ explodes at $0$
if and only if $\beta<1$ (equivalently, $\alpha<1$)
and $2\mu_0<\sigma_0^2$;
$Y$ explodes at $\infty$ if and only if
$\alpha>1$ (equivalently, $\beta>1$)
and $2\mu_0>\sigma_0^2$.
By Theorem~\ref{MRA3},
$Z$ is $\PP$-a.s. strictly positive if and only if
$(\alpha-1)(\sigma_0^2-2\mu_0)\ge0$.

Case~3: $\alpha+1>2\beta$.
Theorem~\ref{MRA3} implies that $Z$ is $\PP$-a.s.
strictly positive if and only if
$Y$ does not explode at~$\infty$.
The latter holds if and only if $\mu_0<0$ or $\alpha\le1$.
Theorem~\ref{MRA4} yields that $Z_\infty>0$~$\PP$-a.s.
if and only if $s(\infty)=\infty$,
and the latter, in turn, holds if and only if $\mu_0<0$.

These findings are summarised in Table~\ref{t:CEV}.
Finally, let us mention that this example complements
Example~3.2 in~\cite{MijatovicUrusov:09a},
where it is studied for which parameter values
$Z$ is a strict local martingale, a martingale,
and a uniformly integrable martingale.

\newlength{\mywidth}
\setlength{\mywidth}{48mm}
\newlength{\mywidtha}
\setlength{\mywidtha}{71mm}
\begin{table}[htb!]
\begin{center}
\scalebox{0.95}{
\begin{tabular}{|c||p{\mywidtha}|p{\mywidth}|}
\hline
Case
&\makebox[\mywidtha]{$Z=(Z_t)_{t\in[0,\infty)}$}
&\makebox[\mywidth]{$Z_\infty$}\\[1mm]
\hline\hline
$\alpha+1<2\beta$
&$Z_t>0\;\;\PP\text{-a.s.}\iff\mu_0>0\;\;\text{or}\;\;\alpha\ge1$
&\makebox[\mywidth]{$Z_\infty=0\;\;\PP\text{-a.s.}$}\\[1mm]
\hline
$\alpha+1=2\beta$
&$Z_t>0\;\;\PP\text{-a.s.}\iff(\alpha-1)(\sigma_0^2-2\mu_0)\ge0$
&\makebox[\mywidth]{$Z_\infty=0\;\;\PP\text{-a.s.}$}\\[1mm]
\hline
$\alpha+1>2\beta$
&$Z_t>0\;\;\PP\text{-a.s.}\iff\mu_0<0\;\;\text{or}\;\;\alpha\le1$
&\parbox{\mywidth}{\vspace*{1mm}
always $\PP(Z_\infty>0)>0$;\\
$Z_\infty>0\;\;\PP\text{-a.s.}\iff \mu_0<0$}\\[4mm]
\hline
\end{tabular}
}
\end{center}
\vspace*{2mm}
\caption{Classification in Example~\ref{ex:EAA2}.}
\label{t:CEV}
\end{table}
\end{example}


\section{Several notions of arbitrage}
\label{SNA}
Let us consider the state space $J=(0,\infty)$ and a $J$-valued diffusion $Y=(Y_t)_{t\in[0,\infty)}$
on some filtered probability space $(\Omega,\cF,(\cF_t)_{t\in[0,\infty)},\PP)$ driven by the SDE
\begin{equation}
\label{sna1}
dY_t=\mu(Y_t)\,dt+\sigma(Y_t)\,dW_t,\quad Y_0=x_0>0,
\end{equation}
where $W$ is an $(\cF_t)$-Brownian motion and $\mu,\sigma$ are Borel functions $J\to\bbR$.
The filtration $(\cF_t)$ is assumed to be right-continuous but we assume neither that $(\cF_t)$ is generated by $W$
nor that $(\cF_t)$ is generated by~$Y$.
The process $Y$ represents the discounted price process of an asset.
In this section we assume the following:

(A) $\sigma(x)\ne0$ $\forall x\in J$;

(B) $1/\sigma^2\in\LlocJ$;

(C) $\mu/\sigma^2\in\LlocJ$;

(D) $Y$ does not explode at $\infty$.

\noindent
Conditions (A)--(C) are the Engelbert--Schmidt conditions
which guarantee the uniqueness in law for SDE~\eqref{sna1}
as well as the existence of a filtered probability space
that supports a (possibly explosive) weak solution of~\eqref{sna1}.
In the case the filtration $(\cF_t)$ is (initially) not right-continuous,
we substitute it with the smallest right-continuous filtration that contains it
(the process $W$ remains a Brownian motion with respect to the new filtration
and $Y$ still solves SDE~\eqref{sna1} after such a transformation).
As before, we assume 
that $Y$ is stopped after the explosion time $\zeta$.
Assumption~(D) for the price 
process $Y$ is quite natural.

In this section we present deterministic criteria
in terms of $\mu$ and $\sigma$
for NFLVR, NGA and NRA
and examine how these notions relate to each other.
As stated above
we assume neither that the filtration
$(\cF_t)$
is generated by 
the solution 
$Y$
of SDE~\eqref{sna1}
nor by
the driving Brownian motion~$W$.
It is therefore interesting to note that 
the deterministic criteria 
for NFLVR, NGA and NRA
we are about to describe,
depend only on the functions
$\mu$
and
$\sigma$
and not on the filtration. 
This implies that in our setting these
notions of arbitrage
are independent of the choice of
a right-continuous filtration
with respect to which
$W$
is a Brownian motion
and 
$Y$
is adapted.

Let us start by introducing the conditions
\begin{gather}
\frac{\mu^2}{\sigma^4}\in\LlocJ,
\label{sna2}\\[1mm]
\frac{x\mu^2(x)}{\sigma^4(x)}\in\LlocZ,
\label{sna3}\\[1mm]
\frac{x}{\sigma^2(x)}\notin\LlocZ,
\label{sna4}
\end{gather}
which will be used below,
and explain their meaning.
A natural candidate for the density of an equivalent martingale measure
is the process $Z$ of~\eqref{mra1} with $b:=-\mu/\sigma$.
Condition~\eqref{sna2} is then just a reformulation of condition~\eqref{mra2}
for the specific choice of 
$b$.
Note that we do not assume in this section that~\eqref{sna2} holds
(only a weaker condition~(C) is assumed).
In the case where \eqref{sna2} does hold, condition~\eqref{sna3}
is satisfied if and only if the boundary point~$0$ is good.
Indeed, the auxiliary diffusion of~\eqref{mra16}
is now driven by the driftless SDE $d\wt Y_t=\sigma(\wt Y_t)\,d\wt W_t$,
hence we can take $\wt\rho\equiv1$ and $\wt s(x)=x$,
which clearly reduces \eqref{mra15} to~\eqref{sna3}.
Finally, condition~\eqref{sna4} holds if and only if
the driftless auxiliary diffusion $\wt Y$ does not 
explode at~$0$ (see~\eqref{mra23}).

\subsection{Free lunch with vanishing risk}
\label{FLVR}
We first recall the definition of NFLVR introduced by
Delbaen and Schachermayer in~\cite{DelbaenSchachermayer:94}.
Let an $\bbR^d$-valued semimartingale
$S=(S_t)_{t\in[0,T]}=(S^1_t,\ldots,S^d_t)_{t\in[0,T]}$
be a model for discounted prices of $d$ assets.
The time horizon $T$ is finite or infinite
and in the case $T=\infty$ we understand $[0,T]$ as $[0,\infty)$.
An $\bbR^d$-valued predictable process
$H=(H_t)_{t\in[0,T]}=(H^1_t,\ldots,H^d_t)_{t\in[0,T]}$
is called a \emph{(trading) strategy} in the model $S$
if the stochastic integral
$(H\cdot S_t)_{t\in[0,T]}:=(\int_0^t H_u\,dS_u)_{t\in[0,T]}$
is well-defined.\footnote{See~\cite[Ch III, Sec. 6c]{JacodShiryaev:03}
for the definition of vector stochastic integrals with semimartingale 
integrators and integrands that are not necessarily locally bounded.}
Here $H^i_t$ is interpreted as the number of assets of type $i$
that an investor holds at time~$t$.
The process $x+H\cdot S$, $x\in\bbR$, is the \emph{(discounted) wealth process}
of the trading strategy $H$ with the initial capital~$x$.
A strategy $H$ is called \emph{admissible}
if there exists a nonnegative constant $c$ such that
\begin{equation}
\label{eq:Admissible}
H\cdot S_t\geq-c\quad\text{a.s.}\quad\forall t\in[0,T].
\end{equation}
Condition~\eqref{eq:Admissible}
rules out economically infeasible risky strategies which attempt
to make a certain final gain by allowing an unbounded amount of
loss in the meantime.
The convex cone of contingent claims attainable from zero initial capital is given by
\begin{equation*}
K:=\{H\cdot S_T\>|\>H\>\text{is admissible and if}\>T=\infty,\>\text{then}\>
H\cdot S_\infty:= \lim_{t\to\infty}H\cdot S_t\>\text{exists a.s.}\}.
\end{equation*}
Let $C$ be the set of essentially bounded random variables
that are dominated by the attainable claims in~$K$.
In other words let
\begin{equation*}
C:=\{g\in L^\infty\>|\>\exists
f\in K\>\>\text{such that}\>\>g\leq f\text{ a.s.}\}.
\end{equation*}
We say that 
the model $S$
satisfies the 
\textit{NFLVR}
condition 
if 
\begin{equation}
\label{eq:NFLVR_closure}
\ol C\cap L^\infty_+=\{0\},
\end{equation}
where 
$\ol C$
denotes the closure of 
$C$
in
$L^\infty$
with respect to the norm topology
and
$L^\infty_+$
denotes the cone of non-negative elements in
$L^\infty$.

The fact that the closure in equation~\eqref{eq:NFLVR_closure}
is in the topology induced by the norm (and not in some weak topology)
has financial significance.
Assume that there is FLVR in the model~$S$.
Then there exists an element
$g\in L_+^\infty\backslash\{0\}$
and a sequence of bounded contingent claims
$(g_n)_{n\in\bbN}$,
which is almost surely dominated 
by
a sequence of attainable claims
$(f_n)_{n\in\bbN}$
in 
$K$
(i.e.
$g_n\leq f_n$
a.s.
and
$f_n=H^n\cdot S_T$
where
$H^n$
is an admissible strategy
for all
$n\in\bbN$),
such that 
\begin{equation*}
\lim_{n\to\infty}\|g-g_n\|_\infty=0,
\end{equation*}
where
$\|\cdot\|_\infty$
is the essential supremum norm on
$L^\infty$.
In particular the sequences
$(f_n\wedge0)_{n\in\bbN}$
and
$(g_n\wedge0)_{n\in\bbN}$
tend to zero uniformly. This implies
that the risks of the admissible trading strategies 
$(H^n)_{n\in\bbN}$,
that correspond to the attainable
claims
$(f_n)_{n\in\bbN}$,
vanish with increasing 
$n$.
It is this interpretation of the definition of
NFLVR that makes it economically meaningful.

The main result in~\cite{DelbaenSchachermayer:98},
which is a generalisation of the main result in~\cite{DelbaenSchachermayer:94},
states that such a model $S$
satisfies NFLVR if and only if there exists an
equivalent sigma-martingale measure for~$S$.
Together with the Ansel-Stricker lemma this implies that
if each component of $S$ is locally bounded from below,
then NFLVR holds if and only if there exists an
equivalent local martingale measure for~$S$.
For further discussions we refer to~\cite{DelbaenSchachermayer:06} and the references therein.

\medskip
In our setting the solution $Y$ of SDE~\eqref{sna1},
which does not explode at $\infty$ but might explode at~$0$,
is a real-valued nonnegative semimartingale and therefore satisfies NFLVR
if and only if there exists a probability measure $\QQ\sim\PP$
such that $(Y_t)_{t\in[0,T]}$ is an $(\cF_t,\QQ)$-local martingale.
We first characterise NFLVR in the model
$Y$ on a finite time horizon.

\begin{theorem}
\label{FLVR1}
Under assumptions (A)--(D)
the market model~\eqref{sna1} satisfies NFLVR on a finite time interval $[0,T]$
if and only if
at least one of the conditions (a)--(b) below is satisfied:

(a) conditions \eqref{sna2} and~\eqref{sna3} hold;

(b) conditions \eqref{sna2} and~\eqref{sna4} hold,
and $Y$ does not explode at~$0$.
\end{theorem}

Let $s$ be the scale function of diffusion~\eqref{sna1}
and $\rho$ the derivative of~$s$ (see \eqref{mra8} and~\eqref{mra9}).

\begin{remark}
Theorem~\ref{FLVR1} generalises one of the results in~\cite{DelbaenShirakawa:02b},
where NFLVR on a finite time interval is characterised under stronger assumptions
using techniques different to the ones employed here.
Namely, in~\cite{DelbaenShirakawa:02b}
the authors work in the canonical setting
(essentially this means that their filtration is generated by~$Y$)
and assume additionally that functions $\mu$, $\sigma$ and $1/\sigma$ 
are locally bounded on~$J$.
In particular in their setting \eqref{sna2} is automatically satisfied.
In this case they obtain that NFLVR holds
if and only if either (a') or~(b') below is satisfied:

(a') \eqref{sna3} holds, \eqref{sna4} is violated, $Y$ explodes at~$0$,
and $\frac{(s-s(0))\mu^2}{\rho\sigma^4}\in\LlocZ$\footnote{Note that
$s(0)>-\infty$ here because $Y$ explodes at~$0$.};

(b') \eqref{sna4} holds and $Y$ does not explode at~$0$.

\noindent
Since the criterion ``(a) or~(b)'' of Theorem~\ref{FLVR1}
looks different from the criterion ``(a') or~(b')'' in~\cite{DelbaenShirakawa:02b},
we need to prove that under~\eqref{sna2} both criteria are equivalent.
We have already observed that under~\eqref{sna2} condition~\eqref{sna3} means that
the endpoint~$0$ is good, i.e. condition~\eqref{sna3} is equivalent to the pair
$s(0)>-\infty$ and $\frac{(s-s(0))\mu^2}{\rho\sigma^4}\in\LlocZ$.
Now the desired equivalence of the two criteria follows from Lemma~\ref{FLVR2} below.
\end{remark}

\begin{lemma}
\label{FLVR2}
Under assumptions (A)--(C) we have the following implication.
If \eqref{sna2} and~\eqref{sna3} hold,
then one of the conditions (i) and (ii) below is satisfied:

(i) \eqref{sna4} holds and $Y$ does not explode at~$0$;

(ii) \eqref{sna4} is violated and $Y$ explodes at~$0$.
\end{lemma}

This lemma is a consequence of remark~(ii) preceding Theorem~\ref{th:MRold}
(see also the discussion following conditions \eqref{sna2}--\eqref{sna4}).

In the case of a non-explosive $Y$ Theorem~\ref{FLVR1} takes the simpler
form of Corollary~\ref{FLVR3}.

\begin{corollary}
\label{FLVR3}
Suppose that (A)--(D) hold and $Y$ does not explode at~$0$.
Then the market model~\eqref{sna1} satisfies NFLVR on a finite time interval $[0,T]$
if and only if
conditions \eqref{sna2} and~\eqref{sna4} are satisfied.
\end{corollary}

The proof follows immediately from Theorem~\ref{FLVR1} and Lemma~\ref{FLVR2}.

Finally, we characterise NFLVR on the infinite time horizon.

\begin{theorem}
\label{FLVR4}
Under assumptions (A)--(D) the market model~\eqref{sna1} satisfies NFLVR on the time interval $[0,\infty)$
if and only if conditions \eqref{sna2} and~\eqref{sna3} hold and $s(\infty)=\infty$.
\end{theorem}

The proofs of Theorems \ref{FLVR1} and~\ref{FLVR4} require
additional concepts and notation 
and are given in Section~\ref{PFLVR}.

\subsection{Generalised arbitrage}
\label{GA}
Sin~\cite{Sin:96} and Yan~\cite{Yan:98} introduced
some strengthenings of NFLVR in continuous time model
with a finite number of assets and a finite time horizon,
and proved that their no-arbitrage notions are
equivalent to the existence of an equivalent
\emph{martingale} measure
(not just a sigma-martingale measure
or a local martingale one).
Later Cherny~\cite{Cherny:07}
introduced the notion of NGA in a certain general setting
including, in particular, continuous time model
with a finite number of assets.
In the latter setting Cherny's characterisation
of NGA coincides with Sin's and Yan's characterisations.
Thus, Sin's and Yan's no-arbitrage notions
may be termed NGA as well.

We first recall the definition of NGA from~\cite{Cherny:07}
and do it only in continuous time model with a finite number of assets.
This will make clear the difference with NFLVR.
Let a model for discounted prices of $d$ assets
be an $\bbR^d$-valued adapted c\`adl\`ag process
$S=(S_t)_{t\in[0,T]}$ with nonnegative components.
The time horizon $T$ is finite or infinite.
In the case $T=\infty$ we understand $[0,T]$ as $[0,\infty)$
and assume\footnote{This assumption is superfluous
but the definition of NGA looks much more technical without it.
On the other hand it turns out that NGA on $[0,\infty)$
does not hold whenever that assumption is violated;
see Section~5 in~\cite{Cherny:07}.
Since we are just recalling the definition of NGA here and want to make it transparent,
it is natural to take that assumption now.
In what follows we will use only a characterisation of NGA on $[0,\infty)$
(Corollary~5.2 in~\cite{Cherny:07}), which applies regardless of
whether that assumption does or does not hold.}
that the limit $S_\infty:=\text{(a.s.)}\,\lim_{t\to\infty}S_t$
exists in~$\bbR^d$.
We consider the set of $\bbR^d$-valued simple
predictable trading strategies $H=(H_t)_{t\in[0,T]}$,
i.e. the processes of the form
\begin{equation}
\label{eq:SimpeStrat}
H=\sum_{k=1}^N h_{k-1} I_{(\tau_{k-1},\tau_k]},
\end{equation}
where $N\in\bbN$, $0\le\tau_0\le\cdots\le\tau_N\le T$
are stopping times, and $h_{k-1}$ are $\bbR^d$-valued
$\cF_{\tau_{k-1}}$-measurable random variables.
Here the set of contingent claims attainable from zero
initial capital is given by
\begin{equation*}
K:=\{H\cdot S_T\>|\>H\text{ is a simple strategy}\},
\end{equation*}
where the ``stochastic integral'' $H\cdot S$ is understood in the obvious way
(in the case $T=\infty$ no problems arise due to our assumption on~$S$).
At this point one can see that short selling in a market model
that is not bounded from above (e.g. the Black--Scholes model)
is allowed~--- something that is not admissible in the context of NFLVR.
Now let
\begin{equation*}
C:=\{h\in L^\infty\>|\>\exists f\in K\text{ such that }
h\le f/Z_0\text{ a.s.}\}
\end{equation*}
with $Z_0:=1+\sum_{i=1}^d S^i_T$
($S^i$ is the $i$-th component of~$S$).
The model $S$ satisfies \emph{NGA} if
\begin{equation}
\label{eq:NGA_closure}
\ol C^*\cap L^\infty_+=\{0\},
\end{equation}
where $\ol C^*$ denotes the closure of $C$
in the topology $\sigma(L^\infty,L^1)$ on~$L^\infty$
(the weak-star topology).

The ramification of the fact that the closure in~\eqref{eq:NGA_closure}
is taken with respect to the weak-star topology and not the topology
induced by the norm on $L^\infty$ is that it might not be possible
to construct a countable sequence of the simple trading strategies~\eqref{eq:SimpeStrat}
that can exploit the existence of generalised arbitrage in the model
(it is of course always possible to find a net, i.e. a generalised sequence,
of elements in $C$ that converge in the weak-star topology to a nonnegative payoff,
strictly positive with a positive probability).
This perhaps makes the notion of GA less economically meaningful.
However the mathematical characterisation of NGA is very transparent.
It is proved in~\cite{Cherny:07} that under the assumptions above
the model $S$ satisfies NGA if and only if there exists an equivalent
probability measure under which the process $S=(S_t)_{t\in[0,T]}$
is a uniformly integrable martingale.\footnote{If $T$ is finite,
this is of course equivalent to the existence of an equivalent
probability measure under which $S$ is a martingale.}
In particular, NGA implies NFLVR.

\medskip
Let us now characterise NGA on a finite time horizon in the model
$Y=(Y_t)_{t\in[0,T]}$ given by SDE~\eqref{sna1}.

\begin{theorem}
\label{GA1}
Under assumptions (A)--(D)
the market model~\eqref{sna1} satisfies NGA on a finite time interval $[0,T]$
if and only if
NFLVR holds on $[0,T]$ \textup{(}see Theorem~\ref{FLVR1} and Corollary~\ref{FLVR3}\textup{)}
and $x/\sigma^2(x)\notin\Llocinfty$.
\end{theorem}

\begin{proof}
1) Suppose that we have NGA on $[0,T]$.
This means that there exists a probability measure $\QQ\sim\PP$
such that $(Y_t)_{t\in[0,T]}$ is an $(\cF_t,\QQ)$-martingale.
Then the process
\begin{equation*}
W'_t:=\int_0^t \frac1{\sigma(Y_s)}\,dY_s
=W_t+\int_0^t \frac\mu\sigma(Y_s)\,ds,
\quad t\in[0,\zeta\wedge T),
\end{equation*}
is a continuous $(\cF_t,\QQ)$-local martingale
on the stochastic interval $[0,\zeta\wedge T)$
with $\langle W',W'\rangle_t=t$, $t\in[0,\zeta\wedge T)$,
hence an $(\cF_t,\QQ)$-Brownian motion stopped at $\zeta\wedge T$.
In other words there exists a Brownian motion~$B$, possibly defined
on an enlargement of the initial probability space,
such that, when stopped at the stopping time
$\zeta\wedge T$,
it satisfies
$B^{\zeta\wedge T}=W'$
(see \cite[Ch.~V, Th.~1.6]{RevuzYor:99}).
Thus, under $\QQ$ the process $(Y_t)_{t\in[0,T]}$ satisfies the SDE
\begin{equation}
\label{ga1}
dY_t=\sigma(Y_t)\,dB_t,\quad Y_0=x_0,
\end{equation}
because by definition the process $(Y_t)_{t\in[0,T]}$
is stopped after the explosion time~$\zeta$.
Since $(Y_t)_{t\in[0,T]}$ is a true martingale under~$\QQ$,
we get $x/\sigma^2(x)\notin\Llocinfty$
by Corollary~4.3 in~\cite{MijatovicUrusov:09a}.
It remains to recall that NGA implies NFLVR.

\smallskip
2) Conversely, assume that NFLVR holds on $[0,T]$
and $x/\sigma^2(x)\notin\Llocinfty$.
Then there exists a probability measure $\QQ\sim\PP$
such that $(Y_t)_{t\in[0,T]}$ is an $(\cF_t,\QQ)$-local martingale.
A similar argument to the one above implies that $(Y_t)_{t\in[0,T]}$
satisfies SDE~\eqref{ga1} under~$\QQ$.
By Corollary~4.3 in~\cite{MijatovicUrusov:09a}, the condition $x/\sigma^2(x)\notin\Llocinfty$
guarantees that $(Y_t)_{t\in[0,T]}$ is an $(\cF_t,\QQ)$-martingale.
This concludes the proof.
\end{proof}

In contrast to the finite time horizon case described by
Theorem~\ref{GA1},
in our setting
GA is always present 
on the infinite time horizon.

\begin{proposition}
\label{GA2}
Under assumptions (A)--(D)
there always exists GA in the market model~\eqref{sna1}
on the time interval $[0,\infty)$.
\end{proposition}

\begin{proof}
Assume that NGA holds.
Then there exists a probability measure $\QQ\sim\PP$
such that $(Y_t)_{t\in[0,\infty)}$ is a uniformly integrable
$(\cF_t,\QQ)$-martingale.
But under $\QQ$ the process $(Y_t)_{t\in[0,\infty)}$
satisfies SDE~\eqref{ga1}, hence, by Corollary~4.3 in~\cite{MijatovicUrusov:09a},
it cannot be a uniformly integrable $(\cF_t,\QQ)$-martingale.
This contradiction concludes the proof.
\end{proof}

\subsection{Arbitrage relative to the market}
\label{RA}
We first recall the definition of relative arbitrage using the terminology and notations
introduced in the beginning of Section~\ref{FLVR}.
The concept of RA appears in the context of stochastic portfolio theory (SPT).
In SPT it is typically assumed that asset prices are strictly positive
It\^o processes. Thus, we consider here a $d$-dimensional It\^o process
$S=(S^1_t,\ldots,S^d_t)_{t\in[0,T]}$ with strictly positive
components as a model for discounted prices of $d$ assets.
The time horizon $T$ is finite. Let $V^{x,H}=(V^{x,H}_t)_{t\in[0,T]}$
denote the (discounted) wealth process of a trading strategy $H$
with the initial capital~$x$, i.e. $V^{x,H}_t=x+H\cdot S_t$.
Here only strategies $(x,H)$ with strictly positive wealth
$V^{x,H}$ will be considered.

\begin{remark}
In the literature on SPT strategies are usually parametrized
in a way different from that above.
Here $H^i_t$ is interpreted as the number of assets of type $i$
that an investor holds at time~$t$;
then the wealth in the money market is determined
automatically by the condition that the strategy
is self-financing.
In the literature on SPT a \emph{strategy} with the initial
capital $v>0$ is $\pi=(\pi^1_t,\ldots,\pi^d_t)_{t\in[0,T]}$,
where $\pi^i_t$ represents the proportion of total wealth
$V^{v,\pi}_t$ invested at time $t$ in the $i$-th asset;
then the proportion of total wealth invested in the money market
at time $t$ is just $1-\sum_{j=1}^d \pi^j_t$
(note that $\pi^i$ and $1-\sum_{j=1}^d \pi^j$ are allowed
to take negative values).
It is easy to check that the set of the strategies $(v,\pi)$
in the latter sense coincides with the set of the strategies
$(x,H)$ with strictly positive wealth.
That is why we prefer not to introduce new notations,
but rather to consider strategies $(x,H)$
as in the beginning of Section~\ref{FLVR}
with strictly positive wealth.
\end{remark}

The \emph{market portfolio} is the strategy
$H^\cM\equiv(1,\ldots,1)$ with the initial capital
$\sum_{i=1}^d S^i_0$, so that its wealth process
$V^\cM$ is given by the formula
$V^\cM=\sum_{i=1}^d S^i_t$.
The terminology becomes clear if we assume that
the stock prices $S^i$, $i=1,\ldots,d$,
are normalized in such a way that each stock
has always just one share outstanding;
then $S^i_t$ is interpreted as the capitalization of the $i$-th
company at time $t$ and $V^\cM_t$ as the total
capitalization of the market at time~$t$.

We now state the definition of RA as given in~\cite{FernholzKaratzas:08}.
There is \emph{arbitrage relative to the market}
(or simply RA) in the model $S$ if there exists a strategy
with a strictly positive wealth process $V$ that beats
the market portfolio, i.e. $V_0=V^\cM_0$,
$V_T\ge V^\cM_T$~a.s., and $\PP(V_T>V^\cM_T)>0$.
Let us finally note that if some strategy $(V^\cM_0,H)$
realises RA in the model $S$, we cannot conclude that
the strategy $(0,H-H^\cM)$ realises FLVR
because the latter strategy may be non-admissible,
i.e. condition~\eqref{eq:Admissible} may be violated.

\medskip
Our goal is to characterise the absence of RA
on a fixed finite time interval $[0,T]$
in the model $Y$ given by SDE~\eqref{sna1}.
Let us note that in our one-dimensional situation
existence of RA means existence of a strategy
with a strictly positive wealth that beats the stock~$Y$.
To put ourselves in the framework of SPT
we suppose that (A), (B), (C') and~(D') hold, where

(C') $\mu^2/\sigma^4\in\LlocJ$;

(D') $Y$ explodes neither at $0$ nor at~$\infty$.

\noindent
As it was mentioned above strictly positive asset prices
are considered in SPT; so we arrive to~(D').
Assumption~(C') is, by the occupation times formula, equivalent to
\begin{equation}
\label{ra1}
\int_0^T \frac{\mu^2}{\sigma^2}(Y_u)\,du<\infty\quad\PP\text{-a.s.},
\end{equation}
and condition~\eqref{ra1} is usually assumed in the literature as well.
For further details see e.g.
\cite{FernholzKaratzas:05},
\cite{FernholzKaratzasKardaras:05},
\cite{FernholzKaratzas:09},
\cite{FernholzKaratzas:08},
and~\cite{Ruf:09}.

Let $\cF^Y_t:=\bigcap_{\eps>0} \sigma(Y_s\,|\,s\in[0,t+\eps])$
be the right-continuous filtration generated by~$Y$.
Let us consider the exponential local martingale
\begin{equation*}
Z_t=\exp\left\{-\int_0^t\frac\mu\sigma(Y_u)\,dW_u
-\frac12\int_0^t\frac{\mu^2}{\sigma^2}(Y_u)\,du\right\},
\end{equation*}
where $W$ is the driving Brownian motion in~\eqref{sna1}.
By It\^o's formula we get that the process $ZY=(Z_tY_t)_{t\in[0,T]}$
is an $(\cF_t)$-local martingale.

\begin{lemma}
\label{RA0}
Under assumptions (A), (B), (C') and~(D')
the market model~\eqref{sna1} satisfies NRA on $[0,T]$
if and only if $ZY$ is an $(\cF_t)$-martingale on~$[0,T]$.
\end{lemma}

\begin{remark}
This statement was first observed by Fernholz and Karatzas
in a different situation (see Section~6 in~\cite{FernholzKaratzas:08}).
To apply their result we need the following representation property:
all $(\cF_t)$-local martingales can be represented
as stochastic integrals with respect to~$W$.
In general the latter property does not hold
in our setting because the filtration $(\cF_t)$
is allowed to be strictly greater than~$(\cF^Y_t)$
(note also that $W$ is adapted to $(\cF^Y_t)$
because $\sigma$ does not vanish);
so we cannot just refer to Section~6 in~\cite{FernholzKaratzas:08}.
However, a part of the proof below will be similar
to the argumentation in~\cite{FernholzKaratzas:08}
(it is needed to make the proof self-contained).
\end{remark}

\begin{proof}
1) At first let us assume that $ZY$ is an $(\cF_t)$-martingale
on $[0,T]$ and take a strategy $(x_0,H)$ with a strictly positive
wealth process $V_t=x_0+\int_0^t H_u\,dY_u$
satisfying $V_T\ge Y_T$ $\PP$-a.s.
(recall that $Y_0=x_0$, so we have also $V_0=Y_0$).
By It\^o's formula the process $ZV$ is an $(\cF_t)$-local martingale
starting from~$x_0$.
As a positive local martingale it is a supermartingale.
We have
\begin{equation*}
x_0\ge\EE Z_TV_T\ge\EE Z_TY_T=x_0,
\end{equation*}
hence $V_T=Y_T$ $\PP$-a.s. Thus, NRA on $[0,T]$ holds.

\smallskip
2) Let us now suppose that the process $ZY$
is not an $(\cF_t)$-martingale on $[0,T]$.
As a positive local martingale it is a supermartingale,
hence $x:=\EE Z_TY_T<x_0$.
Let us consider a strictly positive $(\cF^Y_t)$-martingale
\begin{equation*}
M_t:=\EE(Z_TY_T|\cF^Y_t)
\end{equation*}
and an $(\cF^Y_t)$-local martingale
\begin{equation*}
L_t:=Y_t-x_0-\int_0^t \mu(Y_u)\,du.
\end{equation*}
The latter is an $(\cF^Y_t)$-local martingale
as a continuous $(\cF_t)$-local martingale
adapted to $(\cF^Y_t)$.
Indeed, we can take the sequence
of $(\cF^Y_t)$-stopping times
\begin{equation*}
\tau_n=\inf\{t\in[0,\infty)\colon|L_t|>n\}\quad(\inf\emptyset:=\infty)
\end{equation*}
as a localizing sequence.
By the zero-one law at time~$0$ for diffusion~$Y$
the $\sigma$-field $\cF^Y_0$ is $\PP$-trivial,
hence $M_0=x$ $\PP$-a.s.
It follows from uniqueness in law for~\eqref{sna1}
and the Fundamental Representation Theorem
(see~\cite[Ch.~III, Th.~4.29]{JacodShiryaev:03})
that there exists an $(\cF^Y_t)$-predictable process~$K$,
which is integrable with respect to~$L$, such that
\begin{equation*}
M_t=x+\int_0^t K_u\,dL_u\quad\PP\text{-a.s.}
\end{equation*}
Using It\^o's formula we get after some computations that
\begin{equation*}
\frac{M_t}{Z_t}=x+\int_0^t H_u\,dY_u\quad\PP\text{-a.s.}
\end{equation*}
with
\begin{equation*}
H_u:=\frac{K_u\sigma^2(Y_u)+M_u\mu(Y_u)}{Z_u\sigma^2(Y_u)}.
\end{equation*}
Thus, the strategy $(x,H)$ has the strictly positive wealth process $V^{x,H}=M/Z$
with $V^{x,H}_T=Y_T$ $\PP$-a.s. Since $x<x_0$, the strategy $(x_0,x_0 H/x)$
realises RA on $[0,T]$. This completes the proof.
\end{proof}

Now we can prove a deterministic characterisation of NRA in our setting.

\begin{theorem}
\label{RA1}
Under assumptions (A), (B), (C') and~(D')
the market model~\eqref{sna1} satisfies NRA on $[0,T]$
if and only if $x/\sigma^2(x)\notin\Llocinfty$.
\end{theorem}

\begin{proof}
Due to Lemma~\ref{RA0}
it suffices to show that $ZY$ is a martingale
on $[0,T]$ if and only if $x/\sigma^2(x)\notin\Llocinfty$.
It\^o's formula yields
\begin{equation*}
d(Z_tY_t)=Z_tY_tb(Y_t)\,dW_t
\end{equation*}
with $b(x)=\frac{\sigma(x)}x-\frac{\mu(x)}{\sigma(x)}$, hence
\begin{equation*}
Z_tY_t=x_0\exp\left\{\int_0^t b(Y_u)\,dW_u-\frac12\int_0^t b^2(Y_u)\,du\right\},
\end{equation*}
and we can use Theorem~\ref{th:MRold} to understand
when $ZY$ is a martingale.
Let us note that~(C')
implies condition
\eqref{mra2} for the function
$b$
given above.
The auxiliary diffusion $\wt Y$ evolves in this case 
according to the SDE
\begin{equation*}
d\wt Y_t=\frac{\sigma^2(\wt Y_t)}{\wt Y_t}\,dt+\sigma(\wt Y_t)\,d\wt W_t,
\quad\wt Y_0=x_0.
\end{equation*}
A simple computation yields that we can take
$\wt\rho(x)=\frac1{x^2}$, $\wt s(x)=-\frac1x$, $x\in J=(0,\infty)$.
Since $\wt s(0)=-\infty$, the diffusion $\wt Y$ does not explode at~$0$.
It follows from remark~(ii) preceding Theorem~\ref{th:MRold}
that $\infty$ is a bad point whenever $\wt Y$ explodes at~$\infty$
(recall that $Y$ does not explode at $\infty$ due to assumption~(D')).
Now Theorem~\ref{th:MRold} yields that $ZY$ is a martingale
if and only if $\wt Y$ does not explode at~$\infty$.
By Feller's test, $\wt Y$ does not explode at $\infty$
if and only if $x/\sigma^2(x)\notin\Llocinfty$ (see~\eqref{mra21}).
This concludes the proof.
\end{proof}

\subsection{Comparison}
\label{SNAC}
Here we compare NFLVR, NGA and NRA
in the one-dimensional diffusion setting.
Suppose that (A), (B), (C') and~(D') hold
and consider a finite time horizon $T\in(0,\infty)$
so that all three notions can be defined simultaneously.
From the theorems above we observe

(i) NFLVR $\Longleftrightarrow$ $x/\sigma^2(x)\notin\LlocZ$;

(ii) NRA $\Longleftrightarrow$ $x/\sigma^2(x)\notin\Llocinfty$;

(iii) NGA $\Longleftrightarrow$ NFLVR and NRA.

\noindent
Using (i) and~(ii) we easily construct the following examples
(assumptions (A), (B), (C') and~(D') hold in all of them).

(1) If $\sigma(x)=x$ and $\mu(x)=x$, we have NFLVR and NRA.

(2) If $\sigma(x)=x^2$ and $\mu(x)=x$, we have NFLVR and RA.

(3) If $\sigma(x)=2\sqrt{x}$ and $\mu\equiv d$
with some $d\ge2$, we have FLVR and NRA.\footnote{If $d<2$,
then $Y$ will explode at~0, so assumption~(D') will be violated.}

(3') If $Y$ is a three-dimensional Bessel process
(i.e. $\sigma\equiv1$ and $\mu(x)=1/x$), we get again FLVR and NRA.
This is a well-known example. It first appeared in~\cite{DelbaenSchachermayer:95},
where it was shown that even classical arbitrage exists in this model
(see also Example~3.6 in~\cite{KaratzasKardaras:07},
Example~1 in Section~4.2 of~\cite{FernholzKaratzas:08}
and Example~1 in Section~6 of~\cite{Ruf:09}
for explicit constructions of the arbitrage).

(4) If $\sigma(x)=\sqrt{x}+x^2$ and $\mu\equiv2$, we have FLVR and RA.

We conclude that NFLVR and NRA are in a general position
and their relation to NGA is given in item~(iii) above.

\section{Proofs of the characterisation theorems of Sections~\ref{MR} and~\ref{FLVR}}
\label{PFLVR}
The proofs rely on the notion of separating time for a pair of measures
on a filtered space (see~\cite{ChernyUrusov:04}) and on results on the form
of separating times for the distributions of the solutions of one-dimensional SDEs
(see \cite{ChernyUrusov:04} and~\cite{ChernyUrusov:06}).
Section~5 of~\cite{MijatovicUrusov:09a}
gives a brief description of the properties of separating times
that will be used here.

We start in the setting and notation of Section~\ref{MR}.
Additionally we will need to work with the following canonical setting.
As in Section~\ref{MR} let us consider the state space $J=(l,r)$,
where
$-\infty\leq l<r\leq \infty$, 
and set $\ol J=[l,r]$.
Let $\Omega^*:=\ol C([0,\infty),J)$ be the space
of continuous functions $\omega^*\colon[0,\infty)\to\ol J$
that start inside $J$ and can explode, i.e. there exists $\zeta^*(\omega^*)\in(0,\infty]$
such that $\omega^*(t)\in J$ for $t<\zeta^*(\omega^*)$
and in the case $\zeta^*(\omega^*)<\infty$ we have
either $\omega^*(t)=r$ for $t\ge\zeta^*(\omega^*)$
(hence, also $\lim_{t\uparrow\zeta^*(\omega^*)}\omega^*(t)=r$)
or $\omega^*(t)=l$ for $t\ge\zeta^*(\omega^*)$
(hence, also $\lim_{t\uparrow\zeta^*(\omega^*)}\omega^*(t)=l$).
We denote the coordinate process on $\Omega^*$ by $X^*$
and consider the right-continuous canonical filtration
$\cF^*_t=\bigcap_{\eps>0}\sigma(X^*_s\colon s\in[0,t+\eps])$
and the $\sigma$-field $\cF^*=\bigvee_{t\in[0,\infty)}\cF^*_t$.
Note that the random variable $\zeta^*$ described above
is the explosion time of~$X^*$.
Let the probability measures $\PP^*$ and $\tPP^*$
on $(\Omega^*,\cF^*)$ be the distributions of the solutions
of SDEs~\eqref{mr1} 
and~\eqref{mra16}.
By $S^*$ we denote the separating time for
$(\Omega^*,\cF^*,(\cF^*_t)_{t\in[0,\infty)},\PP^*,\tPP^*)$.
An explicit form of $S^*$ is given in Theorem~5.5 in~\cite{MijatovicUrusov:09a}
and the structure of $S^*$ is described in remark~(ii) following this theorem.

As usual let 
$\PP^*_t$
(resp.
$\tPP^*_t$)
denote the restriction of 
$\PP^*$
(resp.
$\tPP^*$)
to the measurable space
$(\Omega^*,\cF^*_t)$
for any
$t\in[0,\infty]$.
Let the measure
$\tQQ^*_t$
be the absolutely continuous part of
$\tPP^*_t$
with respect to the measure
$\PP^*_t$.

Let
$Z^*$
be the stochastic exponential 
defined on the canonical probability space,
which is analogous to the process
$Z$
given in~\eqref{mra1}.
For the precise definition of
$Z^*$
see~\cite[Sec.~6,~Eq.~(41)]{MijatovicUrusov:09a}.
It is clear from this definition 
that it suffices to prove Theorems~\ref{MRA3},~\ref{MRA4} and~\ref{MRA5}
in the canonical setting.
Recall that by Lemma~6.4 in~\cite{MijatovicUrusov:09a}
we have the following equality
\begin{equation}
\label{eq:lemma64}
Z^*_t=\frac{d\tQQ^*_t}{d\PP^*_t}\quad\PP^*\text{-a.s.},\quad t\in[0,\infty].
\end{equation}
We now proceed to prove the theorems in Section~\ref{MR}.

\begin{proof}[Proof of Theorem~\ref{MRA3}]
The task is to prove that
$Z^*_T>0\;\;\PP^*\text{-a.s.}$
for a fixed
$T\in(0,\infty)$.
By the equality in~\eqref{eq:lemma64}
we have
\begin{equation*}
Z^*_T>0\;\;\PP^*\text{-a.s.}
\Longleftrightarrow
\frac{d\tQQ^*_T}{d\PP^*_T}>0\;\;\PP^*\text{-a.s.}
\Longleftrightarrow
\PP^*_T\ll\tPP^*_T
\Longleftrightarrow
S^*>T\;\;\PP^*\text{-a.s.},
\end{equation*}
where the second equivalence follows from the 
Lebesgue decomposition of 
$\tPP^*_T$
with respect to 
$\PP^*_T$
and the last equivalence is a consequence of 
the definition of the 
separating time
(see the remark after Lemma~5.4 in~\cite[Sec. 5]{MijatovicUrusov:09a}).

In the case 
$\PP^*\ne\tPP^*$, or equivalently 
$\nu_L(b\ne0)>0$
where 
$\nu_L$
is the Lebesgue measure,
Theorem~5.5
in~\cite{MijatovicUrusov:09a}
implies that
$S^*>T\;\;\PP^*\text{-a.s.}$
if and only if the coordinate process
$X^*$
does not explode under
$\PP^*$
at a bad endpoint of~$J$.
In the case $\nu_L(b\ne0)=0$ (i.e. $\PP^*=\tPP^*$)
we have that if 
$l$ (resp.~$r$) 
is bad, then 
$\wt s(l)=-\infty$ (resp. $\wt s(r)=\infty$),
hence $X^*$ does not explode at~$l$ (resp. at~$r$) under~$\tPP^*$.
The two cases together therefore yield the criterion in Theorem~\ref{MRA3}.
\end{proof}

A similar argument, based on the 
equality in~\eqref{eq:lemma64}
for 
$t=\infty$,
implies 
that 
$Z^*_\infty>0\;\;\PP^*\text{-a.s.}$
(resp.
$Z^*_\infty=0\;\;\PP^*\text{-a.s.}$)
if and only if 
$S^*>\infty\;\;\PP^*\text{-a.s.}$
(resp.
$S^*\leq\infty\;\;\PP^*\text{-a.s.}$).
Theorem~5.5
and
Propositions A.1 -- A.3
in~\cite{MijatovicUrusov:09a}
imply Theorems~\ref{MRA4}
and~\ref{MRA5}. The details are very similar to the 
ones in the proof above and are omitted.

\medskip
In order to prove the theorems of Section~\ref{SNA},
we need to recast the canonical space 
$(\Omega^*,\cF^*,(\cF^*_t)_{t\in[0,\infty)},\PP^*,\tPP^*)$
into the setting of that section.
In particular
we take the state space
$J=(0,\infty)$
in the definition of 
$\Omega^*$
and
define 
the probability measures 
$\PP^*$ and $\tPP^*$ 
to be the distributions of the solutions
of the SDEs~\eqref{sna1} and 
$d\wt Y_t=\sigma(\wt Y_t)\,d\wt W_t$, $\wt Y_0=x_0$, respectively.
In all that follows the notation is as in Section~\ref{SNA}.

\begin{proof}[Proof of Theorem~\ref{FLVR1}]
1) Suppose that we have NFLVR on a finite time interval $[0,T]$.
This means that there exists a probability measure $\QQ\sim\PP$
on $(\Omega,\cF)$ such that $(Y_t)_{t\in[0,T]}$ is an $(\cF_t,\QQ)$-local martingale.
Then under $\QQ$ the process $(Y_t)_{t\in[0,T]}$ satisfies the SDE
\begin{equation}
\label{pflvr1}
dY_t=\sigma(Y_t)\,dB_t,\quad Y_0=x_0,
\end{equation}
with some Brownian motion~$B$,
possibly defined on an enlargement of 
the initial probability space
(see the paragraph 
in Section~\ref{SNA}
where~\eqref{ga1} is given
for the precise description of the process~$B$).
Let the probability measure $\QQ^*$ on $(\Omega^*,\cF^*)$
be the distribution of $Y$ with respect to~$\QQ$.
Since $(Y_t)_{t\in[0,T]}$ satisfies~\eqref{pflvr1} under~$\QQ$,
we get $\QQ^*_{T-\eps}=\tPP^*_{T-\eps}$ for any $\eps>0$
($\eps$ appears here due to the fact that $(\cF^*_t)$ is the
right-continuous canonical filtration).
Let us recall that $\PP^*$ is the distribution of $Y$
with respect to~$\PP$.
Since $\QQ\sim\PP$, then
$\tPP^*_{T-\eps}\sim\PP^*_{T-\eps}$ for any $\eps>0$.
By the remark following Lemma~5.4 in~\cite{MijatovicUrusov:09a},
we get $S^*\ge T$ $\PP^*,\tPP^*$-a.s.
Then we need to apply Theorem~5.5 in~\cite{MijatovicUrusov:09a}
and remark~(ii) after it to analyse the implications of
the property $S^*\ge T$~$\PP^*,\tPP^*$-a.s.
We obtain that at least one of the conditions (a)--(b)
in Theorem~\ref{FLVR1} is satisfied.

\smallskip
2) It remains to prove that if at least one of conditions (a)--(b)
in Theorem~\ref{FLVR1} holds, then we have NFLVR on $[0,T]$.
Let us note that pursuing the reasoning above in the opposite
direction would give us NFLVR in the model
$(\Omega^*,\cF^*,(\cF^*_t),\PP^*)$ with the discounted price~$X^*$.
But this does not give us NFLVR in our model
$(\Omega,\cF,(\cF_t),\PP)$ with the discounted price~$Y$
(note that the filtration $(\cF_t)$ need not be generated by~$Y$,
while $(\cF^*_t)$ is the right-continuous filtration generated by~$X^*$).
Therefore, we must follow a different approach.
To this end, below we work directly in the model
$(\Omega,\cF,(\cF_t),\PP)$ of Section~3
and not in the canonical setting.

Let us assume that at least one of conditions (a)--(b)
in Theorem~\ref{FLVR1} holds and consider an $(\cF_t,\PP)$-local martingale
\begin{equation*}
Z_t=\exp\left\{-\int_0^{t\wedge\zeta}\frac\mu\sigma(Y_s)\,dW_s
-\frac12\int_0^{t\wedge\zeta}\frac{\mu^2}{\sigma^2}(Y_s)\,ds\right\},
\quad t\in[0,\infty),
\end{equation*}
where we set $Z_t:=0$ for $t\ge\zeta$
on $\{\zeta<\infty,\int_0^\zeta \frac{\mu^2}{\sigma^2}(Y_s)\,ds=\infty\}$.
This is exactly the process $Z$ in~\eqref{mra1} with $b(x):=-\mu(x)/\sigma(x)$,
and it is well-defined because assumption~\eqref{mra2} is in our case~\eqref{sna2},
which is present in both condition~(a) and condition~(b) of Theorem~\ref{FLVR1}.
We now need to apply Theorems \ref{MRA3} and~\ref{th:MRold}
to the process~$Z$. The auxiliary diffusion $\wt Y$ of~\eqref{mra16}
is in our case given by $d\wt Y_t=\sigma(\wt Y_t)\,d\wt W_t$, $\wt Y_0=x_0$.
Hence we can take $\wt\rho\equiv1$ and $\wt s(x)=x$.
In particular, $\wt s(\infty)=\infty$ and $\wt Y$ does not explode at $\infty$
(see~\eqref{mra21}). By assumption~(D) in Section~\ref{SNA},
$Y$ does not explode at $\infty$.
In the case where condition~(b) in Theorem~\ref{FLVR1} is satisfied
neither  $\wt Y$ nor $Y$ explode at~$0$.
In the case where condition~(a) in Theorem~\ref{FLVR1} holds
the endpoint $0$ is good.
In both cases it follows from Theorems~\ref{MRA3} and~\ref{th:MRold}
that $Z$ is a strictly positive $(\cF_t,\PP)$-martingale.
Hence we can define a probability measure $\QQ\sim\PP$
by setting $\frac{d\QQ}{d\PP}:=Z_T$.
By Girsanov's theorem the process
\begin{equation*}
W'_t:=W_t+\int_0^{t\wedge T\wedge\zeta}\frac\mu\sigma(Y_s)\,ds,\quad t\in[0,\infty)
\end{equation*}
is an $(\cF_t,\QQ)$-Brownian motion.
Clearly, the process $(Y_t)_{t\in[0,T]}$ satisfies
\begin{equation*}
dY_t=\sigma(Y_t)\,dW'_t,\quad Y_0=x_0.
\end{equation*}
Thus, $(Y_t)_{t\in[0,T]}$ is an $(\cF_t,\QQ)$-local martingale.
This implies that we have NFLVR on $[0,T]$.
\end{proof}

The proof of Theorem~\ref{FLVR4} is similar to that of Theorem~\ref{FLVR1}.
To prove the necessity of the condition one again needs to use Theorem~5.5
in~\cite{MijatovicUrusov:09a} and remark~(ii) after it.
To show the sufficiency one applies Theorem~\ref{MRA4} in the present paper
and Theorem~2.3 in~\cite{MijatovicUrusov:09a},
instead of Theorems \ref{MRA3} and~\ref{th:MRold}.
We omit the details.

\bibliographystyle{abbrv}
\bibliography{refs}

\begin{thebibliography}{10}

\bibitem{Cherny:07}
A.~Cherny.
\newblock General arbitrage pricing model: probability approach.
\newblock {\em Lecture Notes in Mathematics}, 1899:415--446, 2007.

\bibitem{ChernyEngelbert:05}
A.~Cherny and H.-J. Engelbert.
\newblock {\em Singular {S}tochastic {D}ifferential {E}quations}, volume 1858
  of {\em Lecture Notes in Mathematics}.
\newblock Springer-Verlag, Berlin, 2005.

\bibitem{ChernyUrusov:04}
A.~Cherny and M.~Urusov.
\newblock Separating times for measures on filtered spaces.
\newblock {\em Theory Probab. Appl.}, 48(2):337--347, 2004.

\bibitem{ChernyUrusov:06}
A.~Cherny and M.~Urusov.
\newblock On the absolute continuity and singularity of measures on filtered
  spaces: separating times.
\newblock In {\em From Stochastic Calculus to Mathematical Finance. In honor of
  Albert Shiryaev's 70th birthday}, pages 125--168. Springer, Berlin, 2006.

\bibitem{DelbaenSchachermayer:94}
F.~Delbaen and W.~Schachermayer.
\newblock A general version of the fundamental theorem of asset pricing.
\newblock {\em Math. Ann.}, 300(3):463--520, 1994.

\bibitem{DelbaenSchachermayer:95}
F.~Delbaen and W.~Schachermayer.
\newblock Arbitrage possibilities in {B}essel processes and their relations to
  local martingales.
\newblock {\em Probab. Theory Related Fields}, 102(3):357--366, 1995.

\bibitem{DelbaenSchachermayer:98}
F.~Delbaen and W.~Schachermayer.
\newblock The fundamental theorem of asset pricing for unbounded stochastic
  processes.
\newblock {\em Math. Ann.}, 312(2):215--250, 1998.

\bibitem{DelbaenSchachermayer:98a}
F.~Delbaen and W.~Schachermayer.
\newblock A simple counter-example to several problems in the theory of asset
  pricing.
\newblock {\em Mathematical Finance}, 8(2):1--11, 1998.

\bibitem{DelbaenSchachermayer:06}
F.~Delbaen and W.~Schachermayer.
\newblock {\em The mathematics of arbitrage}.
\newblock Springer Finance. Springer-Verlag, Berlin, 2006.

\bibitem{DelbaenShirakawa:02b}
F.~Delbaen and H.~Shirakawa.
\newblock No arbitrage condition for positive diffusion price processes.
\newblock {\em Asia-Pacific Financial Markets}, 9:159--168, 2002.

\bibitem{EngelbertSchmidt:85}
H.-J. Engelbert and W.~Schmidt.
\newblock On one-dimensional stochastic differential equations with generalized
  drift.
\newblock In {\em Stochastic {D}ifferential {S}ystems (Marseille-Luminy,
  1984)}, volume~69 of {\em Lecture Notes in Control and Inform. Sci.}, pages
  143--155. Springer, Berlin, 1985.

\bibitem{EngelbertSchmidt:91}
H.-J. Engelbert and W.~Schmidt.
\newblock Strong {M}arkov continuous local martingales and solutions of
  one-dimensional stochastic differential equations. {III}.
\newblock {\em Math. Nachr.}, 151:149--197, 1991.

\bibitem{FernholzKaratzas:08}
D.~Fernholz and I.~Karatzas.
\newblock On optimal arbitrage.
\newblock {\em Preprint, available at:
  \textup{http://www.math.columbia.edu/\~{}ik/OptArb.pdf}}, 2008.

\bibitem{Fernholz:02}
R.~Fernholz.
\newblock {\em Stochastic portfolio theory}, volume~48 of {\em Applications of
  Mathematics (New York)}.
\newblock Springer-Verlag, New York, 2002.
\newblock Stochastic Modelling and Applied Probability.

\bibitem{FernholzKaratzas:05}
R.~Fernholz and I.~Karatzas.
\newblock Relative arbitrage in volatility-stabilized markets.
\newblock {\em Annals of Finance}, 1(2):149--177, 2005.

\bibitem{FernholzKaratzas:09}
R.~Fernholz and I.~Karatzas.
\newblock Stochastic portfolio theory: an overview.
\newblock {\em Handbook of Numerical Analysis}, 15:89--168, 2009.

\bibitem{FernholzKaratzasKardaras:05}
R.~Fernholz, I.~Karatzas, and C.~Kardaras.
\newblock Diversity and relative arbitrage in equity markets.
\newblock {\em Finance Stoch.}, 9(1):1--27, 2005.

\bibitem{JacodShiryaev:03}
J.~Jacod and A.~N. Shiryaev.
\newblock {\em Limit {T}heorems for {S}tochastic {P}rocesses}, volume 288 of
  {\em Grundlehren der Mathematischen Wissenschaften}.
\newblock Springer-Verlag, Berlin, second edition, 2003.

\bibitem{KaratzasKardaras:07}
I.~Karatzas and C.~Kardaras.
\newblock The num\'eraire portfolio in semimartingale financial models.
\newblock {\em Finance Stoch.}, 11(4):447--493, 2007.

\bibitem{KaratzasShreve:91}
I.~Karatzas and S.~E. Shreve.
\newblock {\em Brownian {M}otion and {S}tochastic {C}alculus}, volume 113 of
  {\em Graduate Texts in Mathematics}.
\newblock Springer-Verlag, New York, second edition, 1991.

\bibitem{Lyasoff:08}
A.~Lyasoff.
\newblock The {FTAP} in the special case of {I}t\^o process financial market.
\newblock {\em Preprint, available at:
  \textup{http://andrew.lyasoff.com/FTAP\_Ito.pdf}}, 2008.

\bibitem{MijatovicUrusov:09a}
A.~Mijatovi\'{c} and M.~Urusov.
\newblock On the martingale property of certain local martingales.
\newblock {\em Preprint}, 2009.

\bibitem{PlatenHeath:06}
E.~Platen and D.~Heath.
\newblock {\em A Benchmark Approach to Quantitative Finance}.
\newblock Springer Finance. Springer-Verlag, Berlin, 2006.

\bibitem{RevuzYor:99}
D.~Revuz and M.~Yor.
\newblock {\em Continuous {M}artingales and {B}rownian {M}otion}, volume 293 of
  {\em Grundlehren der Mathematischen Wissenschaften}.
\newblock Springer-Verlag, Berlin, third edition, 1999.

\bibitem{Ruf:09}
J.~Ruf.
\newblock Optimal trading strategies under arbitrage.
\newblock {\em Preprint, available at:
  \textup{http://www.stat.columbia.edu/\~{}ruf/}}, 2009.

\bibitem{Sin:96}
C.~A. Sin.
\newblock {\em Strictly local martingales and hedge ratios in stochastic
  volatility models}.
\newblock Cornell University, Ithaka, NY, 1996.
\newblock PhD thesis.

\bibitem{Yan:98}
J.-A. Yan.
\newblock A new look at the fundamental theorem of asset pricing.
\newblock {\em J. Korean Math. Soc.}, 35(3):659--673, 1998.

\end{thebibliography}
\end{document}